\documentclass{article}

\usepackage{arxiv}

\usepackage[utf8]{inputenc} 
\usepackage[T1]{fontenc}    
\usepackage{hyperref}       
\usepackage{url}            
\usepackage{booktabs}       
\usepackage{amsfonts}       
\usepackage{nicefrac}       
\usepackage{microtype}      
\usepackage{lipsum}
\usepackage{subcaption}
\usepackage{graphicx}
\usepackage{amsmath}
\graphicspath{ {./images/} }

\usepackage{graphicx}
\usepackage{tabularx}
\usepackage{array}
\usepackage{xcolor}

\newcommand{\thumb}[2]{%
  \includegraphics[height=#1, width=\linewidth, keepaspectratio]{#2}%
}

\title{Smart Target Point Control for Gaussian Splatting Methods}

\author{
 Pratik Bisht \\
  Computer Graphics\\
  University of Siegen\\
  \texttt{pratik.bisht@uni-siegen.de} \\
   \And
 Andreas Kolb \\
  Computer Graphics\\
  University of Siegen\\
  \texttt{andreas.kolb@uni-siegen.de} \\
}

\begin{document}
\maketitle
\begin{abstract}
Standard Gaussian splatting methods rely on heuristic densification and pruning to adaptively allocate primitives during training, and the resulting Gaussian count strongly influences both reconstruction quality and runtime. This makes comparisons across methods fragile: improvements can stem from higher representational capacity rather than algorithmic design. A common and naive workaround for this is hard-stopping or budgeting densification/pruning once a target count is reached, which biases training because different methods hit the cap at different times, yielding non-uniform densify/prune exposure across views and uneven point distributions. We propose a target point control scheme that preserves the standard densification window and cadence, but adjusts only the existing densification and opacity-culling hyper-parameters to track a quadratic target count trajectory. This quota-governor reaches the desired count by 15k iterations without abrupt cutoffs, ensuring that all methods and views receive equal densification and pruning cycles, enabling fairer, capacity-matched evaluation.
\end{abstract}

\keywords{Gaussian Splatting \and Novel View Synthesis \and Differentiable Rendering \and Densification and Pruning \and Benchmarking Protocol}

\section{Introduction}
Gaussian Splatting~\cite{kerbl3Dgaussians} has emerged as a strong baseline for real-time novel view synthesis and neural scene reconstruction, offering significantly faster optimization and rendering than NeRF-style~\cite{mildenhall2021nerf}~\cite{muller2022instant} volumetric methods while maintaining high visual fidelity. Modern splatting pipelines represent a scene as a collection of learnable primitives and rely on training-time \emph{densification} and \emph{pruning} heuristics to adaptively allocate these primitives where reconstruction error is high and remove redundant or low-contributing splats. Variants such as 2D Gaussian Splatting~\cite{huang20242d} and surfel-inspired formulations~\cite{dai2024high} further refine the primitive structure, but still share a similar point-churn mechanism driven by heuristic thresholds and periodic scheduling.

A key challenge for benchmarking such methods is that the final number of primitives is not fixed: it emerges from the interaction between scene content, view sampling, and the chosen densify/prune hyper-parameters. Since primitive count directly impacts both quality and efficiency, comparisons between methods can become ambiguous: an approach may report improved PSNR or SSIM simply because it ends training with substantially more Gaussians, while another may appear faster because it converges with fewer primitives but at degraded fidelity. Consequently, attributing gains to algorithmic improvements rather than differences in representational capacity is difficult.

Many works try to control model capacity by limiting the number of Gaussians, but they do so in two distinct ways. Some impose a hard cap, e.g., explicitly capping the maximum number of primitives and effectively preventing further growth once the limit is reached~\cite{yu2024gaussiantalker}~\cite{deng2025improving}. Other approaches enforce a budgeted densification policy~\cite{mallick2024taming}: densification continues, but when the budget is exceeded, only the top-ranked candidates are kept and the rest are discarded~\cite{rota2024revising}. However, often these strategies are motivated primarily by better primitive placement or memory constraints rather than evaluation and comparison fairness. Although both strategies bound capacity, they can still bias training dynamics. Different methods (and runs) reach or saturate the budget at different times and under different intermediate point distributions. This can freeze under-reconstructed regions that would normally receive densification later, while already over-reconstructed regions may have consumed the available budget. Because densification and pruning operate on a fixed cadence (e.g., every 100 iterations), budget saturation also yields unequal effective exposure to point-churn across views and time, increasing variance and producing non-uniform point distributions that affect quality comparisons.

In this paper, we propose a minimal, training-compatible alternative: \emph{target point control} for Gaussian splatting. Instead of modifying the underlying heuristics or changing the densification/pruning cadence, we keep the standard schedule intact and adjust only the existing densification gradient threshold and opacity-culling threshold so that the duration of densification-pruning cycles stay the same and all cameras/view-points get the same preference. We define a target primitive-count trajectory $N^{\ast}(t)$ over the canonical densification window (up to 15k iterations in common settings) and use a lightweight quota-governor that updates these thresholds smoothly in log-space under bounded step sizes. We adopt a quadratic fast-start trajectory, which allocates capacity early and improves robustness to late-stage disturbances such as opacity resets, enabling the method to reach the approximate desired budget at the end of the densification window without abrupt cutoffs.

Our goal is not to improve reconstruction quality directly, but to provide a fairer evaluation protocol: methods can be compared under matched capacity while preserving the original point-churn behavior of classical splatting pipelines. We recommend reporting both unconstrained results and controlled-budget results using the proposed schedule to separate algorithmic improvements from capacity effects.

\paragraph{Contributions.}
\begin{itemize}
    \item We identify and analyze why hard budget cutoffs in Gaussian splatting bias training dynamics, leading to non-optimal point distributions, making cross-method comparisons unreliable.
    \item We introduce a target point control scheme that preserves the standard densification/pruning cadence and modifies only pre-existing thresholds to track a quadratic target count trajectory.
\end{itemize}

\section{Related Work}
\label{s:related}

\subsection{Gaussian Splatting and Adaptive Density Control}
3D Gaussian Splatting (3DGS) represents scenes as a set of anisotropic Gaussian primitives optimized from multi-view images and rendered via fast differentiable rasterization~\cite{kerbl3Dgaussians}. A core component of 3DGS training is \emph{Adaptive Density Control} (ADC), which grows the set of primitives through densification (e.g., splitting/cloning based on position gradient-driven criteria) and removes redundant/unused primitives via opacity-based pruning. This point-churn mechanism is critical for reconstruction quality and efficiency, but is also heuristic and sensitive to hyperparameters such as densification thresholds, pruning thresholds, cadence, and opacity resets.

Recent work has revisited ADC to improve the stability and effectiveness of densification and pruning. For example, several methods propose improved candidate selection signals (e.g., using additional cues beyond 2D position gradients)~\cite{kim2024color}, refine when-to-densify logic, or introduce significance-aware pruning to mitigate redundancy and background artifacts while keeping the overall training framework intact~\cite{deng2025improving}. These approaches primarily target better reconstruction quality per primitive or better allocation of primitives in textured versus low-frequency regions, but they typically do not address the benchmarking question of enforcing a comparable primitive budget across methods.

\subsection{Compact and Budgeted Gaussian Splatting}
A parallel line of work focuses on \emph{compact} 3DGS representations through pruning strategies, compression, or structured training schedules. More recent methods explicitly incorporate \emph{primitive budgeting} or memory constraints into training, in order to control maximum memory growth from densification and avoid uncontrolled early expansion~\cite{mallick2024taming}~\cite{rota2024revising}. These methods demonstrate that controlling primitive growth is practically important, but their objectives are typically deployment-oriented (e.g., memory-bounded training or compression) rather than providing a general \emph{fair comparison protocol} that preserves a fixed densification/pruning window and cadence across diverse GS variants.

\subsection{Evaluation Protocols and Capacity-Matched Comparisons}
Reproducible evaluation protocols have been a recurring issue in novel view synthesis. Community efforts such as standardized NeRF evaluation frameworks highlight how inconsistent settings (e.g., different training-time budgets, hyperparameter sweeps, or data processing) can obscure scientific conclusions~\cite{kulhanek2024nerfbaselines}. For Gaussian Splatting, comparisons are often additionally confounded by the final number of primitives: higher capacity can yield better metrics, while smaller models may appear faster but underfit. Some papers attempt to control capacity by fixing a maximum number of primitives or reporting results at a chosen budget, but in practice, this strategy can unintentionally bias the spatial allocation of primitives because methods reach the cap at different times and therefore undergo different effective numbers of densify/prune cycles.

\subsection{Our Position}
In contrast to approaches that redesign ADC, introduce new candidate scoring signals, or enforce budgets via abrupt stopping or memory-centric schedules, our goal is purely \emph{evaluation fairness}. We preserve the canonical densification/pruning window and cadence used in standard GS pipelines and adjust only existing densification and pruning thresholds to track a prescribed target primitive trajectory. This yields capacity-matched comparisons without changing the underlying point-churn heuristics, thereby reducing confounding factors when comparing the quality of different splatting methods.

\section{Methodology}
\label{s:method}

Our goal is to enable \emph{capacity-matched} comparisons between Gaussian splatting methods without altering the underlying densification/pruning heuristics or the training schedule. We therefore treat the standard splatting pipeline (rendering, photometric loss, densification, pruning, opacity reset) as a black box and only modulate the existing densification and pruning thresholds such that the total number of primitives reaches a desired target at the end of the canonical densification window (typically 15k iterations). Importantly, we \textbf{do not} change (i) the densification/pruning cadence, (ii) the start/stop iterations of densification, or (iii) the opacity reset schedule. This preserves the original point churn dynamics and avoids spatial biases that arise from prematurely disabling point churn.

Let the scene be represented by a set of $N(t)$ Gaussian primitives at training iteration $t$,
\begin{equation}
\mathcal{G}(t)=\left\{\left(\mu_i,\Sigma_i,c_i,\alpha_i\right)\right\}_{i=1}^{N(t)}.
\end{equation}
In standard 3DGS-style pipelines, the point population changes through a scheduled \emph{densify-and-prune} operator invoked every $c$ iterations (densification cadence, usually 100 iterations). At each invocation, the method (i) selects candidate primitives based on a signal that correlates with reconstruction deficiency, commonly a running statistic of view-space position gradient magnitude, and (ii) creates new primitives by cloning/splitting selected candidates. In parallel, it removes primitives with persistently low opacity via an opacity cull threshold. We denote the two primary hyperparameters as:
\begin{itemize}
    \item Densification threshold $\tau_{\text{den}}$: controls which candidates are densified (lower $\tau_{\text{den}}$ increases the number of candidates passing the criterion, hence increasing point growth).
    \item Pruning threshold $\tau_{\text{prune}}$: controls which primitives are removed (higher $\tau_{\text{prune}}$ prunes more aggressively, as more primitives fall below the opacity cutoff).
\end{itemize}

We denote the method's default thresholds by $\tau_{\text{den},0}$ and $\tau_{\text{prune},0}$.
To prevent degenerate regimes, we constrain controller-updated thresholds to remain within fixed
multiples of these defaults:
\begin{equation}
\label{eq:tau_bounds}
\tau_{\text{den}} \in [\tau_{\text{den,min}}, \tau_{\text{den,max}}],\quad
\tau_{\text{prune}} \in [\tau_{\text{prune,min}}, \tau_{\text{prune,max}}],
\end{equation}

In addition, most pipelines periodically reset opacities (e.g., every $t_{\text{reset}}$ iterations) to escape local minima and encourage rediscovery of structure. This reset is known to transiently disrupt pruning signals and can cause sudden drops if pruning is not handled carefully.

\subsection{Target Point Control: Overview}
\label{subsec:point_control_overview}

We propose \emph{Target Point Control (TPC)}, a lightweight controller that guides the primitive population size to meet $N(t_{\text{stop}})\approx K$ at the end of the densification window, while preserving the original training cadence and heuristics. Let the densification window be $t\in[t_{\text{start}},t_{\text{stop}}]$, where $t_{\text{start}}$ corresponds to \texttt{densify\_from\_iter} and $t_{\text{stop}}$ corresponds to \texttt{densify\_until\_iter}. The densify/prune operator is invoked every $c$ iterations (\texttt{densification\_interval}). We define a target schedule $N^{\ast}(t)$ that specifies the desired number of primitives at every iteration $t$, and update $\tau_{\text{den}}$ and $\tau_{\text{prune}}$ \emph{only} at the same cadence as densify/prune, so that the controller aligns with the discrete actuation structure of the pipeline.

\subsection{Quadratic Target Trajectory}
\label{subsec:quadratic_nstar}

We define a fast-start quadratic schedule to allocate capacity early and reduce late-stage catch-up pressure, especially important when opacity resets. Let $N_0$ be the initial population size at the beginning of training (or at $t_{\text{start}}$), and let $K$ be the final desired population at $t_{\text{stop}}$. For $t\in[t_{\text{start}},t_{\text{stop}}]$, we define
\begin{equation}
\label{eq:nstar_quad}
N^{\ast}(t)=N_0 + s(x)\,(K-N_0), \qquad
x=\frac{t-t_{\text{start}}}{t_{\text{stop}}-t_{\text{start}}}\in[0,1],
\end{equation}
with quadratic fast-start easing
\begin{equation}
\label{eq:quad_ease}
s(x)=1-(1-x)^2 = 2x-x^2.
\end{equation}

We define the target count piecewise as
\begin{equation}
N^{\ast}(t)=
\begin{cases}
N_0, & t \le t_{\text{start}},\\
N_0 + s(x)\,(K-N_0), & t_{\text{start}} < t < t_{\text{stop}},\\
K, & t \ge t_{\text{stop}}.
\end{cases}
\end{equation}
Compared to linear or slow-start schedules (e.g., $x^2$), Eq.~\eqref{eq:quad_ease}
allocates a larger fraction of the total budget earlier (e.g., $s(0.5)=0.75$), which
empirically stabilizes target tracking under opacity resets and reduces the risk of being
far below target near the end of the window.

\begin{figure*}[t]
    \centering
    \begin{subfigure}[t]{0.49\textwidth}
        \centering
        \includegraphics[width=\linewidth]{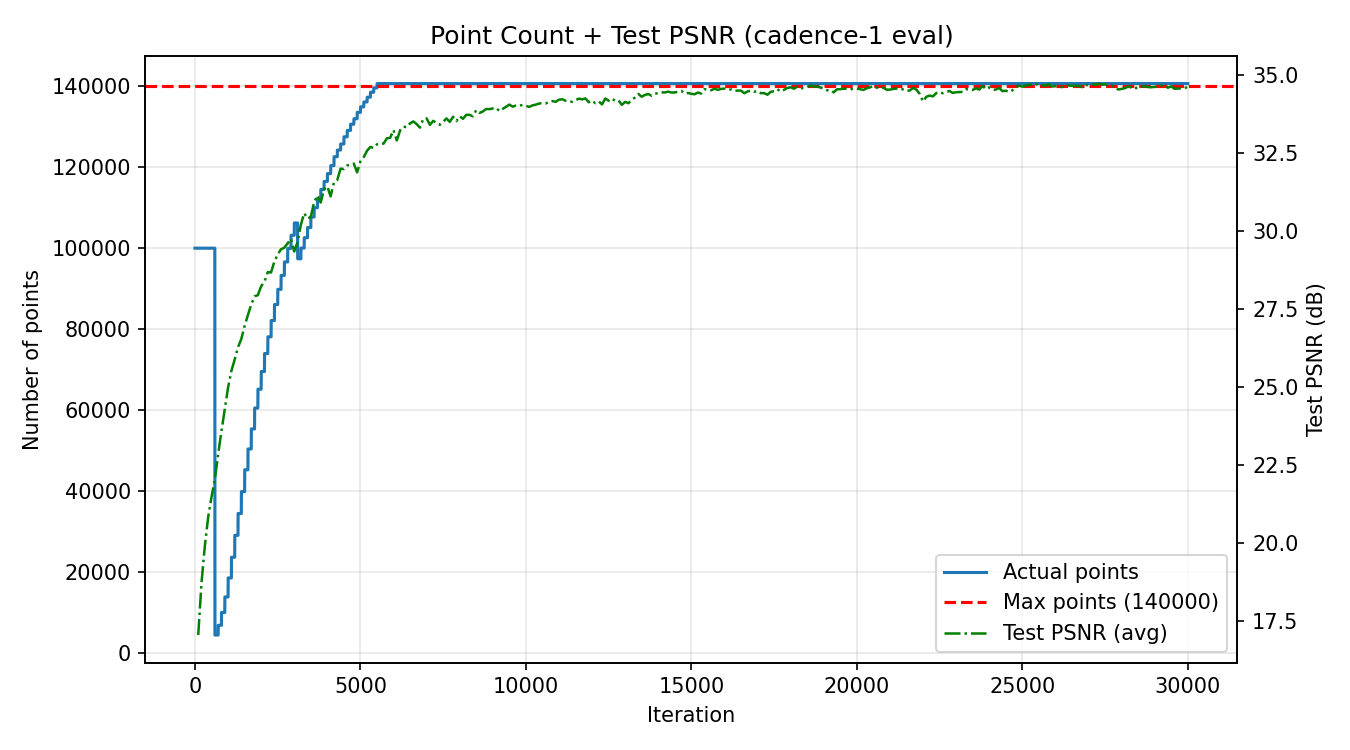}
        \caption{\textbf{Hard cutoff.} Point count is capped once the maximum is reached. The right axis shows the test PSNR trajectory evaluated at \texttt{cadence-1}.}
        \label{fig:lego_cutoff}
    \end{subfigure}\hfill
    \begin{subfigure}[t]{0.49\textwidth}
        \centering
        \includegraphics[width=\linewidth]{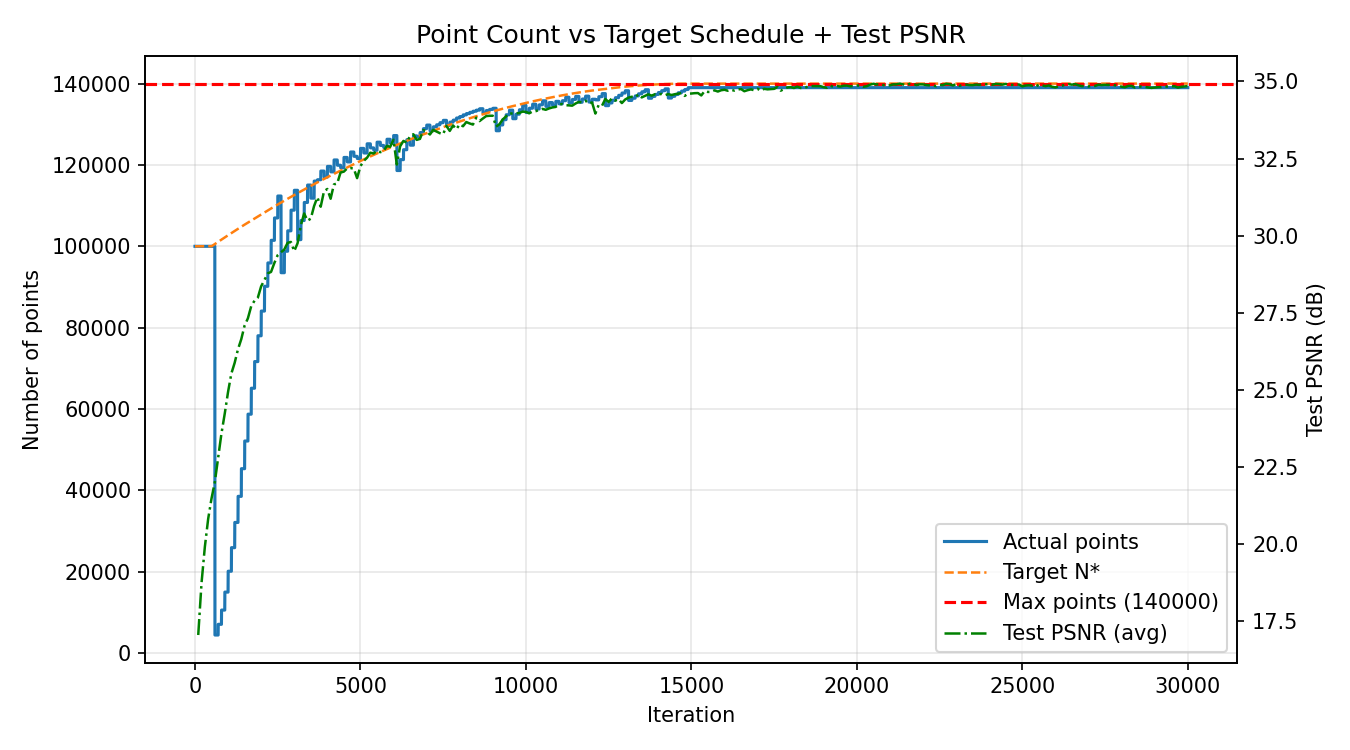}
        \caption{\textbf{TPC (ours).} Point count tracks the target trajectory $N^{\ast}(t)$ during the densification window while reporting the same \texttt{cadence-1} test PSNR curve.}
        \label{fig:lego_tpc}
    \end{subfigure}
    \caption{\textbf{Point growth and test PSNR over training.} Left: hard cutoff. Right: Target Point Control (TPC), which additionally visualizes the target schedule $N^{\ast}(t)$. Both plots show point count on the left axis and test PSNR on the right axis for the Lego scene of NeRF Synthetic dataset.}
    \label{fig:lego_points_psnr_cutoff_vs_tpc}
\end{figure*}

\subsection{Quota-Governor: Discrete Target Tracking}
\label{subsec:quota_governor}

At iteration $t$, let $N(t)$ be the current number of primitives and define the gap
\begin{equation}
\label{eq:gap}
g(t)=N^{\ast}(t)-N(t).
\end{equation}
Since densify/prune is only applied every $c$ iterations, the controller computes how many actuation opportunities remain until $t_{\text{stop}}$:
\begin{equation}
\label{eq:acts_left}
A(t)=1+\left\lfloor\frac{t_{\text{stop}}-t}{c}\right\rfloor,
\quad \text{for } t\le t_{\text{stop}} \text{ aligned to cadence}.
\end{equation}
We then convert the remaining global gap into a per-actuation quota:
\begin{equation}
\label{eq:quota}
q(t)=\text{round}\!\left(\frac{g(t)}{A(t)}\right).
\end{equation}
Intuitively, $q(t)$ specifies how many primitives we should add (if $q(t)>0$) or remove (if $q(t)<0$) per remaining densify/prune call, assuming future calls behave similarly. To avoid chattering when close to the target, we apply:
\begin{itemize}
    \item \textbf{Deadband:} if $|g(t)| < \delta(t)$, set $q(t)=0$, where $\delta(t)$ is proportional to $N^{\ast}(t)$ (e.g., $1\%$) and lower bounded by a constant.
    \item \textbf{Quota floor:} if $|q(t)| < q_{\min}$, set $q(t)=0$, where $q_{\min}$ is set proportional to $K$ (e.g., $K/100$) and optionally increased near the end of the window to enforce decisive corrections.
\end{itemize}

\paragraph{Observed actuation effect.}
We measure the realized population change over the last interval as
\begin{equation}
\label{eq:dn}
\Delta N(t)=N(t)-N(t-1),
\end{equation}
and compare it to the quota. This provides a simple feedback signal that captures the net effect of densification and pruning under current thresholds.

\subsection{Threshold Updates (Bounded Multiplicative Control)}
\label{subsec:log_updates}

We update the densification and pruning thresholds via small \emph{multiplicative} steps,
which are scale-invariant and preserve positivity:
\begin{equation}
\tau \leftarrow \tau\,\exp(\Delta), \qquad |\Delta|\le\Delta_{\max}.
\end{equation}

After each update, we clamp the thresholds to fixed multiples of their default values
$\tau_{\text{den},0}$ and $\tau_{\text{prune},0}$, as defined in \eqref{eq:tau_bounds}.
The bound $\Delta_{\max}$ limits the maximum relative change per update (e.g.,
$\Delta_{\max}=0.12$ corresponds to at most $\approx 12.7\%$ change).

\noindent\textbf{Controller gains and rate limits.}
$\Delta_{\max}>0$ is the maximum magnitude of the log-step applied in one update, limiting the per-update multiplicative change to at most $\exp(\Delta_{\max})$.
$\beta_{\text{den}}>0$ and $\beta_{\text{pru}}>0$ are scalar gains that map the quota tracking errors $e_u(t)$ and $e_o(t)$ to the proposed log-steps $\Delta_{\text{den}}(t)$ and $\Delta_{\text{pru}}(t)$, respectively.

\paragraph{Under-target case ($g(t)>\delta(t)$).}
When the current count is below the target, we minimize pruning by setting the effective pruning threshold
$\tau_{\text{prune}}^{\text{eff}} \leftarrow \tau_{\text{prune,min}}$.
We then adapt the effective densification threshold $\tau_{\text{den}}^{\text{eff}}$ to achieve the required per-step quota.
With $e_u(t)=q(t)-\Delta N(t)$, we apply
\begin{equation}
\Delta_{\text{den}}(t)=-\beta_{\text{den}}\,e_u(t), \qquad
\tau_{\text{den}}^{\text{eff}} \leftarrow \tau_{\text{den}}\,\exp\!\big(\text{clamp}(\Delta_{\text{den}}(t),-\Delta_{\max},\Delta_{\max})\big),
\end{equation}
followed by clamping $\tau_{\text{den}}^{\text{eff}}$ to the allowable range in \eqref{eq:tau_bounds}.
If we added fewer points than desired ($\Delta N(t)<q(t)$), then $e_u(t)>0$ and $\Delta_{\text{den}}<0$,
which decreases $\tau_{\text{den}}$ and increases densification.

\paragraph{Over-target case ($g(t)<-\delta(t)$).}
When the current count is above the target, we prevent further growth by setting the effective densification threshold
$\tau_{\text{den}}^{\text{eff}} \leftarrow \tau_{\text{den,max}}$.
We then adapt pruning to meet the required removal quota.
With $e_o(t)=(-q(t))-(-\Delta N(t))$, we apply
\begin{equation}
\Delta_{\text{pru}}(t)=+\beta_{\text{pru}}\,e_o(t), \qquad
\tau_{\text{prune}}^{\text{eff}} \leftarrow \tau_{\text{prune}}\,\exp\!\big(\text{clamp}(\Delta_{\text{pru}}(t),-\Delta_{\max},\Delta_{\max})\big),
\end{equation}
followed by clamping $\tau_{\text{prune}}^{\text{eff}}$ to the allowable range in \eqref{eq:tau_bounds}.
If we remove fewer points than required, then $e_o(t)>0$ and $\Delta_{\text{pru}}>0$, increasing
$\tau_{\text{prune}}$ and making pruning more aggressive.

\paragraph{Deadband case ($|g(t)|\le\delta(t)$).}
Within the deadband, we keep the thresholds unchanged, i.e.,
$\tau_{\text{den}}^{\text{eff}} \leftarrow \tau_{\text{den}}$ and
$\tau_{\text{prune}}^{\text{eff}} \leftarrow \tau_{\text{prune}}$.
This avoids oscillations when $N(t)$ is already close to the target schedule.

\subsection{Opacity Reset Handling (Prune Lockout)}
\label{subsec:lockout}

The classic pipeline uses the Opacity reset step multiple times during training to bring the system out of local minima. This works well for re-distribution of the primitives to more efficient spatial locations. However, opacity resets can temporarily reduce many $\alpha_i$ values and cause a transient pruning shock, which is required to avoid local minima, but can have an adverse effect on the target point control strategy. Hence we include a prune lockout stage in our approach, which skips the pruning threshold re-tuning based on our controllers and lets the system get back to usual point density naturally before turning on the control knobs. Let $t_{\text{reset}}$ denote an iteration at which opacity resets, and let $T_{\text{lock}}$ be the lockout duration in iterations. To avoid catastrophic drops near the target, we enforce a short pruning lockout after each scheduled opacity reset for $T_{\text{lock}}$ iterations:
\begin{equation}
\tau_{\text{prune}}^{\text{eff}} \leftarrow \tau_{\text{prune,min}}, \quad \text{for } t\in[t_{\text{reset}}, t_{\text{reset}}+T_{\text{lock}}).
\end{equation}
During lockout, densification continues under the standard cadence and the controller can still modulate $\tau_{\text{den}}$ when under-target, enabling recovery without changing the pipeline schedule.

\begin{table*}[t]
\centering
\caption{Unconstrained results for two Gaussian splatting baselines (We report average metrics across scenes for each dataset.)}
\label{tab:combined_mip360_nerfsyn_test}
\resizebox{\textwidth}{!}{
\begin{tabular}{l|ccccc|ccccc}
\toprule
& \multicolumn{5}{c|}{\textbf{Mip-NeRF 360 (Test, Avg.)}} & \multicolumn{5}{c}{\textbf{NeRF-Synthetic (Test, Avg.)}} \\
\midrule
\textbf{Method} &
\textbf{MS-SSIM} $\uparrow$ & \textbf{LPIPS} $\downarrow$ & \textbf{L1} $\downarrow$ & \textbf{PSNR} $\uparrow$ & \textbf{Pts (M)} $\downarrow$ &
\textbf{MS-SSIM} $\uparrow$ & \textbf{LPIPS} $\downarrow$ & \textbf{L1} $\downarrow$ & \textbf{PSNR} $\uparrow$ & \textbf{Pts (M)} $\downarrow$ \\
\midrule
2DGS & 0.894 & 0.074 & 0.021 & 30.01 & 0.833
      & 0.965 & 0.030 & 0.006 & 32.92 & 0.096 \\
\midrule
3DGS & 0.903 & 0.062 & 0.020 & 30.47 & 1.582
      & 0.969 & 0.023 & 0.006 & 33.88 & 0.288 \\
\bottomrule
\end{tabular}}
\end{table*}

\begin{table*}[t]
\centering
\caption{Mip-NeRF 360 capacity-matched results for \textbf{hard cutoff} vs.\ \textbf{TPC}. Each subtable compares 2DGS and 3DGS at the same target point average budget of \textbf{0.785M} points.}
\label{tab:mip360_capacity_test}
\begin{minipage}[t]{0.49\textwidth}
\centering
\subcaption{\textbf{Hard cutoff}}
\label{tab:mip360_cutoff_test}
\resizebox{\textwidth}{!}{
\begin{tabular}{lcccc}
\toprule
\textbf{Method} & \textbf{MS-SSIM} $\uparrow$ & \textbf{LPIPS} $\downarrow$ & \textbf{L1} $\downarrow$ & \textbf{PSNR} $\uparrow$\\
\midrule
2DGS & 0.892 & 0.077 & 0.022 & 29.96 \\
3DGS & 0.897 & 0.073 & 0.021 & 30.33 \\
\bottomrule
\end{tabular}}
\end{minipage}\hfill
\begin{minipage}[t]{0.49\textwidth}
\centering
\subcaption{\textbf{TPC (ours)}}
\label{tab:mip360_tpc_test}
\resizebox{\textwidth}{!}{
\begin{tabular}{lcccc}
\toprule
\textbf{Method} & \textbf{MS-SSIM} $\uparrow$ & \textbf{LPIPS} $\downarrow$ & \textbf{L1} $\downarrow$ & \textbf{PSNR} $\uparrow$\\
\midrule
2DGS & 0.893 & 0.076 & 0.0215 & 30.054 \\
3DGS & 0.901 & 0.068 & 0.0205 & 30.447 \\
\bottomrule
\end{tabular}}
\end{minipage}
\end{table*}

\begin{table*}[t]
\centering
\caption{NeRF-Synthetic capacity-matched results for \textbf{hard cutoff} vs.\ \textbf{TPC}. Each subtable compares 2DGS and 3DGS at the same target point average budget of \textbf{0.0912M} points.}
\label{tab:nerfsyn_capacity_test}
\begin{minipage}[t]{0.49\textwidth}
\centering
\subcaption{\textbf{Hard cutoff}}
\label{tab:nerfsyn_cutoff_test}
\resizebox{\textwidth}{!}{
\begin{tabular}{lcccc}
\toprule
\textbf{Method} &
\textbf{MS-SSIM} $\uparrow$ &
\textbf{LPIPS} $\downarrow$ &
\textbf{L1} $\downarrow$ &
\textbf{PSNR} $\uparrow$ \\
\midrule
2DGS & 0.949 & 0.056 & 0.008 & 30.84 \\
3DGS & 0.952 & 0.054 & 0.008 & 31.32 \\
\bottomrule
\end{tabular}}
\end{minipage}\hfill
\begin{minipage}[t]{0.49\textwidth}
\centering
\subcaption{\textbf{TPC (ours)}}
\label{tab:nerfsyn_tpc_test}
\resizebox{\textwidth}{!}{
\begin{tabular}{lcccc}
\toprule
\textbf{Method} &
\textbf{MS-SSIM} $\uparrow$ &
\textbf{LPIPS} $\downarrow$ &
\textbf{L1} $\downarrow$ &
\textbf{PSNR} $\uparrow$ \\
\midrule
2DGS & 0.966 & 0.030 & 0.006 & 32.94 \\
3DGS & 0.967 & 0.027 & 0.006 & 33.50 \\
\bottomrule
\end{tabular}}
\end{minipage}
\end{table*}

\section{Experiments and Results}
\label{s:experiments}

\subsection{Experimental Setup}
\label{s:setup}

We evaluate \emph{Target Point Control (TPC)} as a \emph{capacity-matching} mechanism for fair comparisons between Gaussian splatting pipelines. The key constraint is that we preserve the original training schedule and heuristics: the \emph{densify-and-prune} operator remains active only within the canonical densification window (ending at \texttt{densify\_until\_iter}, typically 15k iterations) and is invoked at the original cadence \texttt{densification\_interval} (typically 100 iterations). We also keep the opacity reset schedule unchanged. TPC modifies only the \emph{rate} of point churn through the existing densification and pruning thresholds, so all views/cameras experience the same number of densify/prune opportunities.

We consider two representative pipelines: (i) \textbf{3DGS}, and (ii) \textbf{2DGS}. We report results on two datasets:
\textbf{Mip-NeRF 360} (seven scenes: \textit{Bicycle, Bonsai, Counter, Garden, Kitchen, Room, Stump}) and
\textbf{NeRF-Synthetic} (eight scenes: \textit{Chair, Drums, Ficus, Hotdog, Lego, Materials, Mic, Ship}). All the experiments are run on a single RTX 3090 (24GB) with a 24-core CPU and 64GB RAM.

For each pipeline and dataset, we compare three regimes:
\begin{enumerate}
    \item \textbf{No point control (default).} Each method runs with its original densify/prune hyperparameters. This reflects common practice but yields method-dependent final primitive counts.
    \item \textbf{Hard cutoff (abrupt cap).} Densification/pruning is disabled once the point count reaches a fixed target budget. This enforces capacity matching but can terminate point churn early and unevenly across methods and views.
    \item \textbf{TPC (ours).} We preserve the densification window and cadence and only modulate the existing densification and opacity-culling thresholds to track a target point trajectory, reaching the same target budget at the end of the densification window.
\end{enumerate}

\begin{figure*}[t]
\centering
\setlength{\tabcolsep}{3pt} 
\renewcommand{\arraystretch}{1.15}

\newcommand{\imgh}{3.2cm}
\newcommand{\textrise}{0.9cm}

\begin{tabularx}{\textwidth}{>{\raggedright\arraybackslash}p{1.6cm} >{\raggedright\arraybackslash}p{1.8cm} *{3}{>{\centering\arraybackslash}X}}
\toprule
\textbf{Scene} & \textbf{Method} & \textbf{GT} & \textbf{Cutoff} & \textbf{TPC} \\
\midrule

\raisebox{\textrise}{\textbf{Bicycle}} &
\raisebox{\textrise}{3DGS} &
\thumb{\imgh}{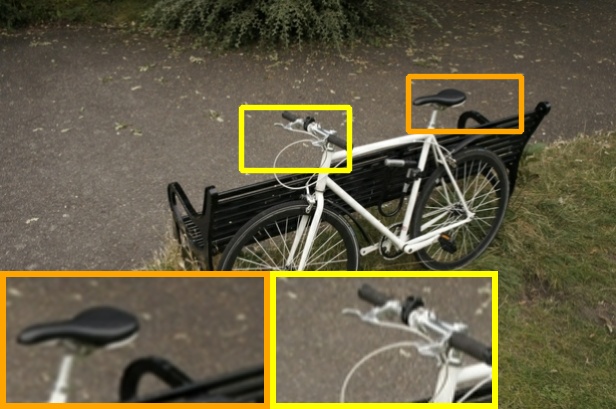} &
\thumb{\imgh}{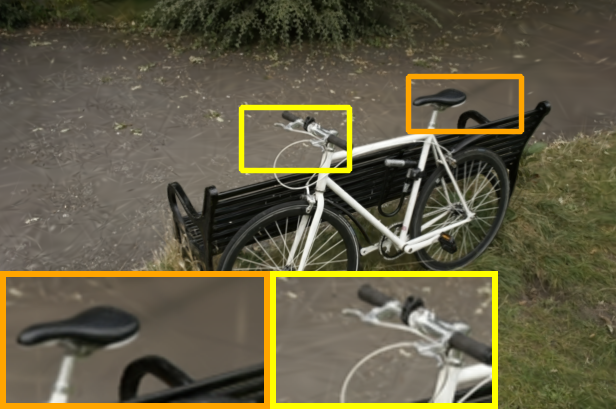} &
\thumb{\imgh}{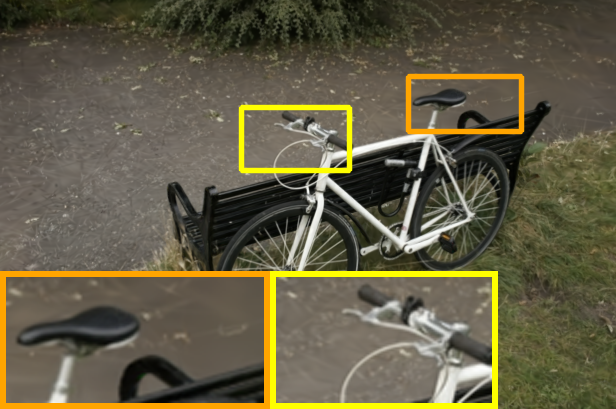} \\

\raisebox{\textrise}{\textbf{Stump}} &
\raisebox{\textrise}{3DGS} &
\thumb{\imgh}{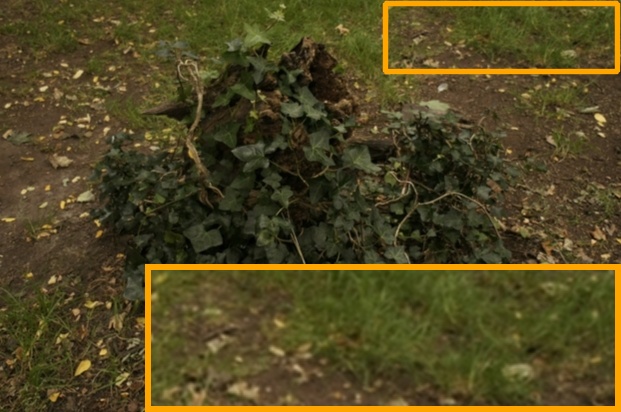} &
\thumb{\imgh}{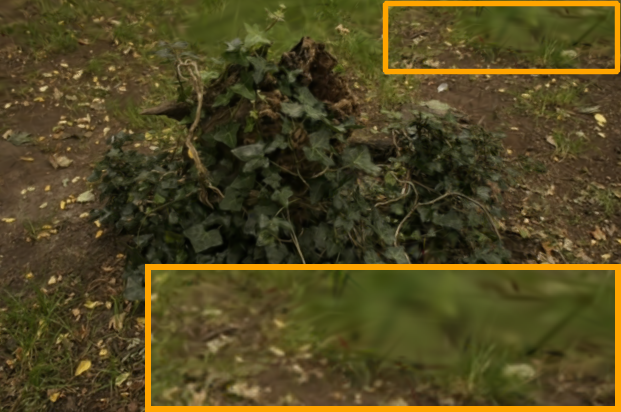} &
\thumb{\imgh}{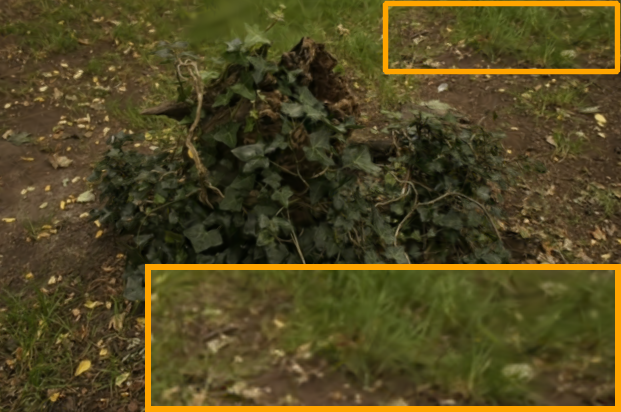} \\

\raisebox{\textrise}{\textbf{Garden}} &
\raisebox{\textrise}{2DGS} &
\thumb{\imgh}{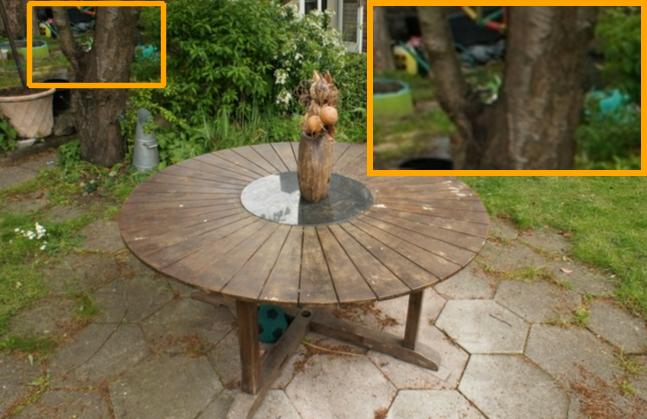} &
\thumb{\imgh}{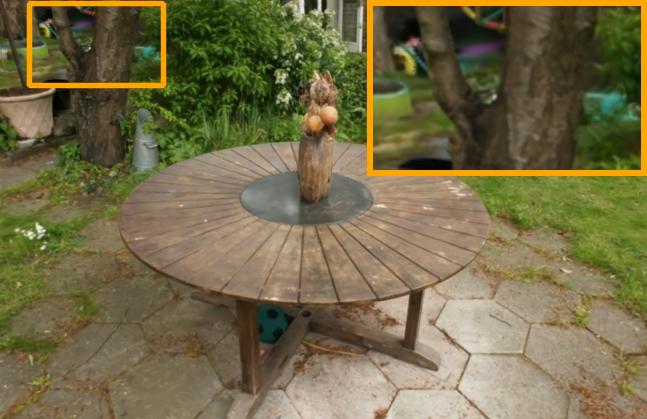} &
\thumb{\imgh}{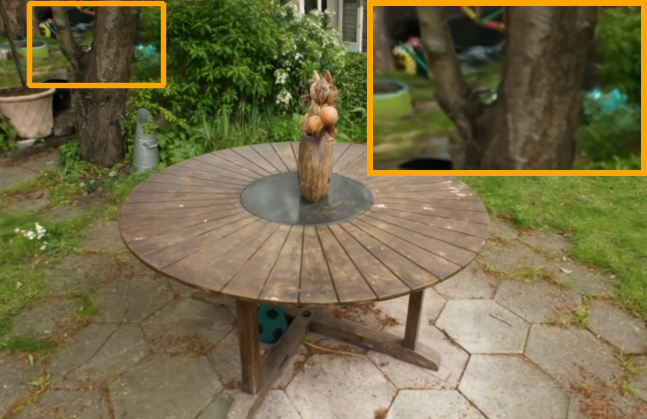} \\

\raisebox{\textrise}{\textbf{Kitchen}} &
\raisebox{\textrise}{2DGS} &
\thumb{\imgh}{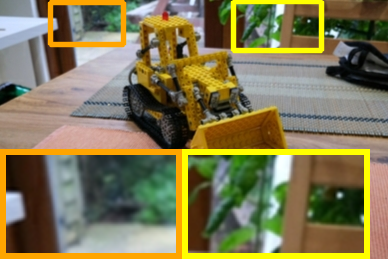} &
\thumb{\imgh}{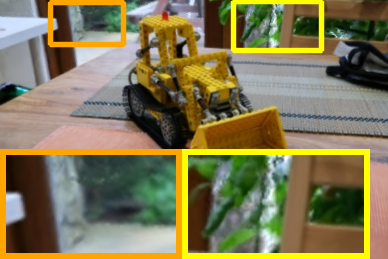} &
\thumb{\imgh}{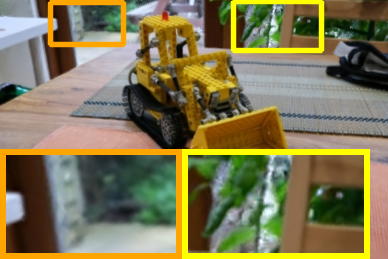} \\

\raisebox{\textrise}{\textbf{Room}} &
\raisebox{\textrise}{2DGS} &
\thumb{\imgh}{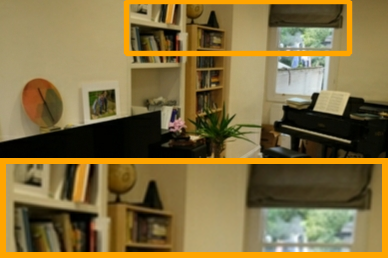} &
\thumb{\imgh}{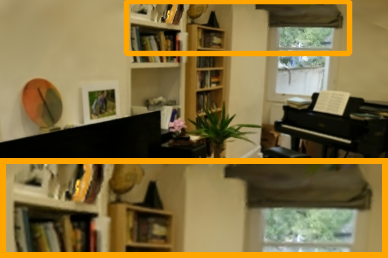} &
\thumb{\imgh}{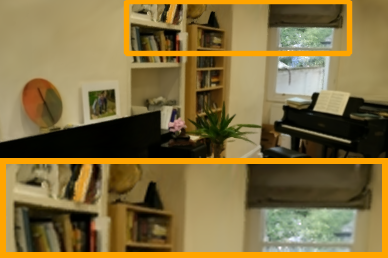} \\

\raisebox{\textrise}{\textbf{Hotdog}} &
\raisebox{\textrise}{3DGS} &
\thumb{\imgh}{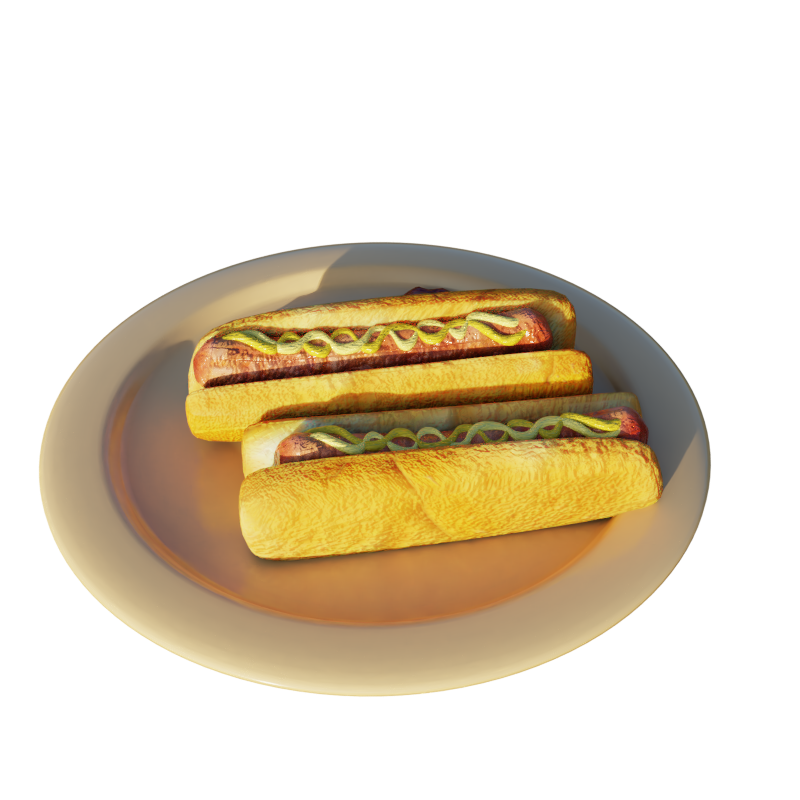} &
\thumb{\imgh}{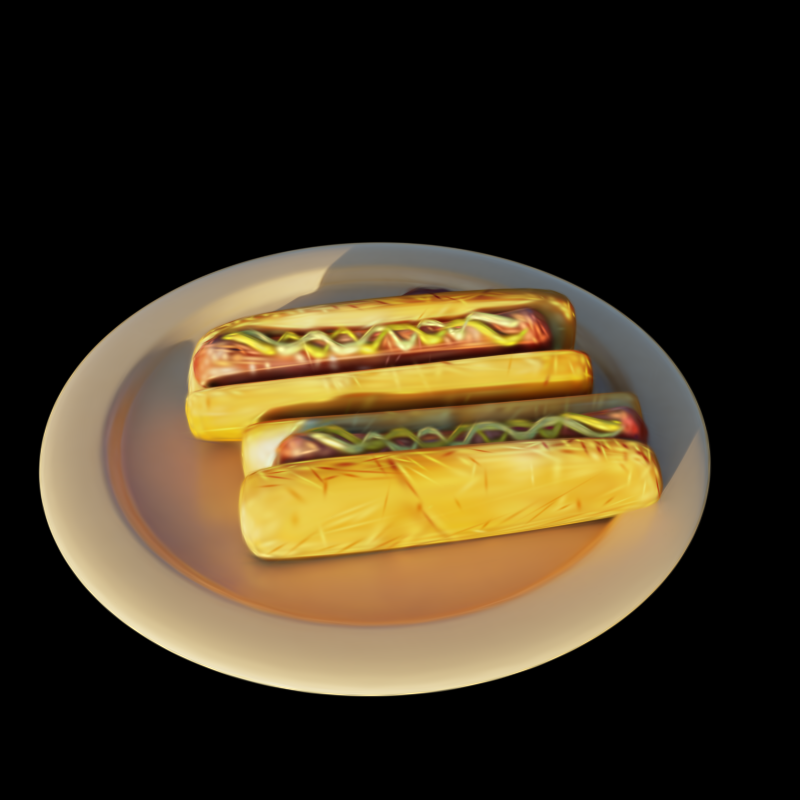} &
\thumb{\imgh}{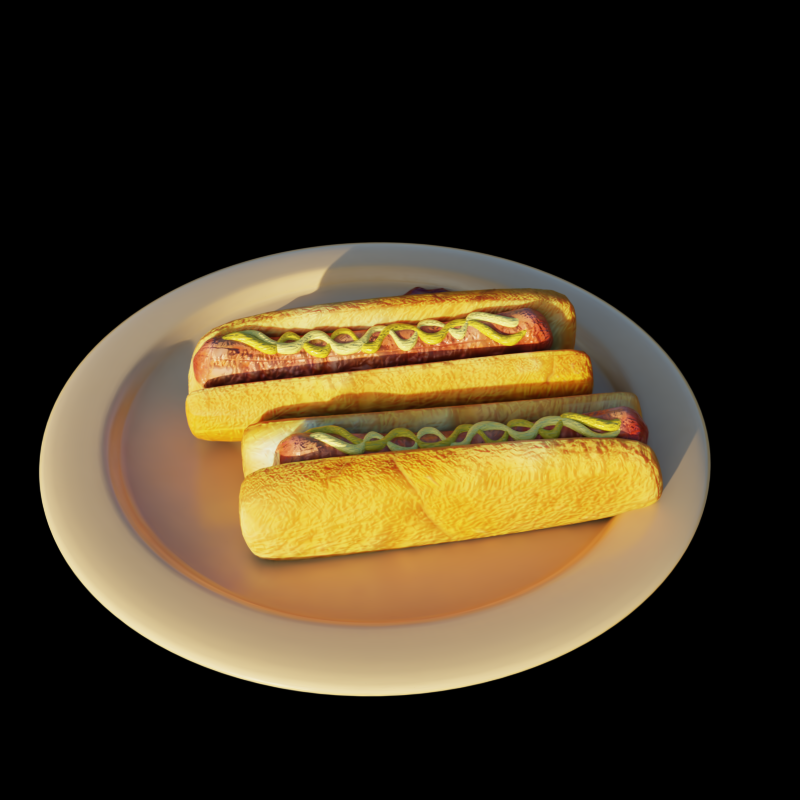} \\

\bottomrule
\end{tabularx}

\caption{\textbf{Qualitative comparison under equal point budgets.} Each row shows a scene/method pair and comparison of Ground truth against cutoff method vs Target Point Control (TPC) method.}
\label{fig:qual_grid_mip_blender}
\end{figure*}

\subsection{Capacity-matched Comparison (Mip-NeRF 360 and NeRF-Synthetic)}
\label{s:capacity_comparison}

A key motivation for this work is that Gaussian splatting pipelines can differ substantially in their \emph{emergent} primitive counts under default densification/pruning, which directly affects both runtime and reconstruction quality. This makes cross-method comparisons fragile: apparent improvements may reflect increased representational capacity rather than algorithmic merit.

\paragraph{Unconstrained training is capacity-confounded.}
Table~\ref{tab:combined_mip360_nerfsyn_test} shows that, without any point control, 3DGS consistently converges to a much larger primitive set than 2DGS across both datasets (e.g., on Mip-NeRF 360: 1.582M vs.\ 0.833M on average; on NeRF-Synthetic: 0.288M vs.\ 0.096M). Consequently, direct comparisons of PSNR/LPIPS/MS-SSIM in this regime are not capacity-matched and can over-attribute gains to a method when they are partly due to a larger primitive budget.

\paragraph{Hard cutoff matches the final budget but changes the optimization trajectory.}
A common capacity-matching strategy is to impose a hard cap by disabling densification (and often pruning) once the primitive count reaches a specified target. While this produces similar final point counts, the time at which each method hits the cap differs, which changes the number of effective densify/prune cycles each method experiences. Because densification and pruning operate on a fixed cadence, hitting the cap early can freeze the primitive distribution before under-reconstructed regions have received their intended share of point churn, whereas over-reconstructed regions may already have consumed the available budget. This makes the final allocation (and thus quality) dependent on when the cap is reached, not only on the underlying method.

\paragraph{TPC preserves canonical point churn while matching the target budget.}
In contrast, Target Point Control (TPC) preserves the canonical densification window, cadence, and opacity reset schedule, and adjusts only the existing densification and pruning thresholds to \emph{reach the same target budget at the end of the window}. This retains the original “exposure” to densify/prune cycles for all methods and avoids the discontinuity introduced by an early cutoff. Under this controlled setting, we compare hard cutoff vs.\ TPC at matched target budgets on both datasets (Mip-NeRF 360 in Table~\ref{tab:mip360_capacity_test}, NeRF-Synthetic in Table~\ref{tab:nerfsyn_capacity_test}). Overall, TPC yields better capacity-matched test performance than hard cutoff, with particularly consistent improvements in perceptual quality (LPIPS) and small but repeatable gains in PSNR/SSIM, indicating that preserving the point-churn dynamics while steering the final count is a more faithful and stable way to enforce fairness.

\paragraph{Qualitative interpretation via point-growth and PSNR curves.}
The figure ~\ref{fig:lego_points_psnr_cutoff_vs_tpc} shows the comparison of point-growth/PSNR between the regimes: hard cutoff typically produces an early plateau in point count (and a corresponding change in the PSNR slope), whereas TPC continues controlled point churn until the end of the densification window, leading to smoother capacity allocation over time and better quality. This helps explain why, at the same final primitive budget, TPC can achieve better test fidelity: it avoids prematurely freezing the distribution and thus reduces the chance that capacity is locked into suboptimal regions or viewpoints. Further visual comparisons are shown in figure ~\ref{fig:qual_grid_mip_blender} for both Mip-NeRF 360 and Blender datasets.

\subsection{Summary and Discussion}
\label{s:summary_discussion}

Across both datasets, the default regime demonstrates that 3DGS and 2DGS naturally converge to very different primitive counts, which makes unconstrained comparisons capacity-confounded. The hard cutoff regime enforces capacity matching but can degrade performance due to altered training dynamics: early termination of point churn changes how many densify/prune opportunities each method and view receives, and it can freeze suboptimal intermediate allocations. TPC achieves the same capacity target while preserving the canonical densification window and cadence, leading to improved test-set metrics compared to hard cutoff in our experiments. The accompanying point-growth and PSNR plots further illustrate the mechanism: cutoff produces an early plateau in point count that can hinder late-stage refinement, whereas TPC continues scheduled churn and converges smoothly to the target, enabling fairer capacity-matched evaluation.

\section{Conclusions}
\label{sec:conclusions}

We studied the problem of \emph{capacity matching} in Gaussian splatting: different pipelines (e.g., 2DGS vs.\ 3DGS) naturally converge to very different primitive counts under default densification/pruning settings, making direct quality comparisons ambiguous. A common workaround is an abrupt hard cutoff that disables point churn once a target budget is reached. While this matches capacity, it also changes the training dynamics in a method-dependent way by prematurely terminating densify/prune cycles, which can bias the spatial allocation of primitives and affect final quality.

To address this, we proposed \emph{Target Point Control (TPC)}, a lightweight controller that preserves the canonical densification window, cadence, and opacity reset schedule, while adjusting only the existing densification and opacity-culling thresholds to track a target primitive-count trajectory and reach the desired budget at the end of the window. Across both Mip-NeRF 360 and NeRF-Synthetic, TPC enables capacity-matched evaluation without abrupt termination and yields improved test-set performance compared to hard cutoff at the same target budgets. 

These results suggest that fair benchmarking for Gaussian splatting should treat the primitive budget as a \emph{controlled variable} rather than an emergent outcome or a late-stage cap. In future work, we plan to explore alternative target trajectories and controller designs that further reduce run-to-run variance while keeping the underlying point-churn heuristics intact.

\bibliographystyle{unsrt}  
\bibliography{references}  

\end{document}